\documentclass[preprint2]{aastex}

\usepackage{fullpage}
\usepackage{longtable}
\usepackage{amssymb}  
\usepackage{multirow}  
\usepackage{amsmath}  
\usepackage{mathrsfs} 
\usepackage{paralist} 
\usepackage[usenames]{color}  
\DeclareGraphicsExtensions{.eps}

\newif{\ifrgb}
\rgbtrue

\newcommand{\TNSN}{TN\,SN}
\newcommand{\TNSNe}{TN\,SNe}
\newcommand{\CCSN}{CC\,SN}
\newcommand{\CCSNe}{CC\,SNe}
\newcommand{\SNuV}{\ensuremath{\mbox{10$^{-4}$~SNIa~Mpc$^{-3}$~yr$^{-1}$~$h_{70}^{3}$}}}

\newcommand{\Av}{{\ensuremath{\mbox{A$_{V}$}}}}
\newcommand{\Rv}{{\ensuremath{\mbox{R$_{V}$}}}}

\newcommand{\tpk}{{\ensuremath{\mbox{t{$_{\mbox{\tiny pk}}$}}}}}
\newcommand{\mue}{\ensuremath{\mu_{e}}}

\newcommand{\Om}{\ensuremath{\Omega_{\mbox{\scriptsize M}}}}
\newcommand{\Ol}{\ensuremath{\Omega_{\scriptsize \Lambda}}}

\defcitealias{Barris:2006}{BT06}
\newcommand{\LCDM}{\ensuremath{\Lambda{\rm CDM}}}

\shorttitle{IfA Deep SN Rates}
\shortauthors{Rodney and Tonry}

\begin{document}
\title{Revised Supernova Rates from the IfA Deep Survey}

\author{Steven A. Rodney}
\affil{Department of Physics and Astronomy,Johns Hopkins University\\
 Baltimore, MD 21218, USA}
\email{rodney@jhu.edu}
\author{John L. Tonry}
\affil{Institute for Astronomy, University of Hawaii\\
Honolulu, HI 96822, USA}
\email{jt@ifa.hawaii.edu}

\begin{abstract}
The IfA Deep survey uncovered $\sim$130 thermonuclear supernovae
(\TNSNe, i.e. Type Ia) candidates at redshifts from $z=0.1$ out to
beyond $z=1$. The \TNSN\ explosion rates derived from these data have
been controversial, conflicting with evidence emerging from other
surveys.  This work revisits the IfA Deep survey to re-evaluate the
photometric evidence.  Applying the SOFT program to the light curves
of all SN candidates, we derive new classification grades and redshift
estimates.  We find a volumetric rate for $z\sim 0.5$ that is
substantially smaller than the originally published values, bringing
the revised IfA Deep rate into good agreement with other surveys.
With our improved photometric analysis techniques, we are able to
confidently extend the rate measurements to higher redshifts, and we
find a steadily increasing \TNSN\ rate, with no indication of a peak
out to $z=1.05$.
\vspace{6mm}
\end{abstract}

\section{Introduction}

In 2001, the IfA Deep Survey was undertaken as a collaborative program
of astronomers at the University of Hawaii's Institute for
Astronomy. The survey was designed with a strong emphasis on
discovering high redshift ($z>0.5$) SNe.  Over a 5 month span the
survey produced 23 spectroscopically confirmed high-z thermonuclear
supernovae (\TNSNe, or Type Ia), doubling the sample at $z>0.7$ and
providing an important verification of the evidence for dark energy
\citep{Barris:2004a}.  In addition, a larger sample of 133
photometrically classified SNe were used to measure the high-z
\TNSN\ explosion rate \citep[][hereafter BT06]{Barris:2006}.
Subsequent measurements from more recent surveys have consistently
found lower values for the \TNSN\ rate at $z\sim 0.5$ than the
\citetalias{Barris:2006}\ results \citep{Neill:2006,Botticella:2008,Kuznetsova:2008}.  The
common explanation for this disagreement has been that the \citetalias{Barris:2006}\ rate
is affected by systematic errors in the photometric classification and
redshift estimation -- which are not present in the spectroscopic
samples of later surveys.

Resolving this discrepancy is an important step toward preparing for
the next generation of very large scale SN surveys.  Spectroscopic
classification and redshift determination are becoming unrealistic for
all-sky survey programs yielding thousands of SNe per year. In order
to make full use of these new kilo-SN data sets we must have a high
confidence in our photometric analysis methods so that truly divergent
results are not mistakenly ascribed to systematic biases.

This work describes a comprehensive reanalysis of the IfA Deep Survey
data using Supernove Ontology with Fuzzy Templates (SOFT), a
fuzzy-set-based method for SN classification and redshift estimation
that does not require any spectroscopic information
\citep{Rodney:2009,Rodney:2010}.  The end result of this analysis is a
recalculation of the \TNSN\ rate as a function of redshift, extending
it out to z=1.1.

\section{The 2006 Rate Estimate}

The IfA Deep Survey was the first survey to take full advantage of the
wide field imaging cameras of Mauna Kea to allow for both SN discovery
and simultaneous follow-up observations. Using the 12k camera on the
Canada France Hawaii Telescope (CFHT) and Suprime-Cam on Subaru, the
survey covered 5 fields with a total survey area of 2.5 deg$^2$.  Each
field was revisited approximately every 10 days for observations in
the R, I and Z bands. The survey design and operations are described
in greater detail by \citet{Barris:2004a} and \citet{Liu:2002}.

The most promising SN candidates were quickly observed
spectroscopically to confirm them as SNe and to measure their
redshifts.  This produced the spectroscopic sample of 23
\TNSNe\ presented by \citet{Barris:2004a}.  After the conclusion of
the 5 month survey, the images were reprocessed to include all
available photometric data in a single analysis. The NN2 algorithm
\citep{Barris:2005} was used to construct light curves from difference
images for over 10,000 possible transient objects.  \citetalias{Barris:2006}\ applied a
succession of screening criteria to identify 133 objects as possible
\TNSN\ candidates.~\footnote{ Note that in \citet{Barris:2004b} and
  \citetalias{Barris:2006}\ there are actually only 131 unique objects, as two SNe were
  counted twice due to a mistake in tabulation. } This list was
reduced to 98 using the original Bayesian Adaptive Template Matching
code (a precursor to the SOFT program used in this work).  It was this
final collection of 98 photometrically classified \TNSNe\ that was
used to determine the IfA Deep SN rates.

\citetalias{Barris:2006}\ recognized that the IfA Deep rate estimates were not in
line with other published measurements.  In particular, the rate at
z=0.55 is discrepant by more than 3$\sigma$ when compared to the
z=0.55 rate of \citet{Pain:2002}.  Subsequently, the principal
critique of the \citetalias{Barris:2006}\ rate estimate has been the possibility of a
systematic bias due to misclassification of core collapse (CC) SNe as
\TNSN.  This point was emphasized by \citet{Neill:2006}, presenting
the \TNSN\ rate measurements of the SNLS project.  They found a rate
at z=0.5 that is $>5\sigma$ less than the \citetalias{Barris:2006}\ rate measurement, and
argued that the most likely source of the discrepancy is low-redshift
CC SNe that were misidentified as \TNSN.

\section{SOFT Reanalysis}
 
The evolution of the light curve analysis code from BATM in BT06 to
SOFT in this work has introduced three significant improvements:
\begin{itemize}
\item The BATM program used a set of 20 \TNSN\ light curve templates
  with no CC templates, SOFT uses 26 \TNSNe\ and 16 CC SNe.
\item BATM uses a purely Bayesian formalism, while SOFT uses fuzzy set
  theory to account for the incompleteness of the set of available
  templates.
\item The BATM code as applied in BT06 provided a binary
  classification (Ia or not) and assigned a single redshift to each
  object. SOFT gives a classification probability and a redshift
  probability distribution.
\end{itemize}
These advances give SOFT much greater leverage for distinguishing
\TNSNe\ from \CCSNe\ \citep{Rodney:2009} and for estimating the redshift
from the light curve alone \citep{Rodney:2010}.

Applying SOFT to each of the 131 SN candidate light curves published
in \citet{Barris:2004b} yields a classification grade 
$\gamma_{\rm TN}$, which approximately reflects the likelihood that each
candidate belongs in that class.\footnote{
  Also called the Posterior Membership Grade, PMG$_{\rm TN}$}  
For priors on the location parameters (time of peak \tpk, host galaxy
extinction \Av, redshift $z$, and distance modulus \mue), we used the
same ``uninformative priors'' that were applied in the validation sets
in \citet{Rodney:2009} -- except when a redshift measurement for the
SN or its host is available.

For the 23 \TNSNe\ from \citet{Barris:2004a}, which have precise
spectroscopic redshift measurements, we applied the SOFT analysis
twice.  The first pass used only the weak, uninformative prior, while
the second pass utilized a much stronger spectroscopic redshift
prior. This subset provides us with a useful internal consistency
check, to validate the SOFT redshift estimates of the remaining 98
SNe.  Table~\ref{tab:hiz23} reports the results of the first pass,
ignoring the spectroscopic information. The classification grade
$\gamma_{\rm TN}$ is reported for each of the 23 spectroscopically
confirmed \TNSNe, along with the measured spectroscopic redshift and
the SOFT redshift estimate when SOFT uses the uninformative $z$ prior.

\begin{figure}[tb!]
  \centering
  \ifrgb{
    \includegraphics[draft=False,width=\columnwidth]{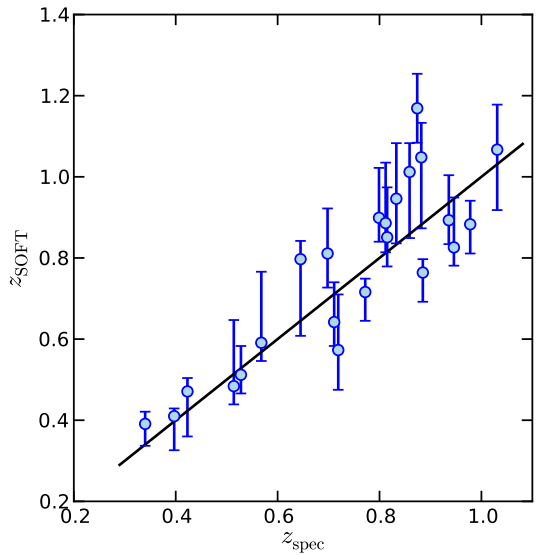} 
  } \else {
    \includegraphics[draft=False,width=\columnwidth]{f01_gray} 
  } \fi
  \caption[SOFT redshift estimates vs. spectroscopic redshifts]{  
    Comparison of the peak redshift estimates from SOFT against
    the spectroscopic redshift measurements for 23 \TNSNe.  
    The straight line shows perfect parity (z$_{SOFT}$=z$_{spec}$).
    The rms scatter about the line is 0.10, and the reduced $\chi^2$  
    statistic is 0.85. 
  }
  \label{fig:zplot23}
\end{figure}

It is encouraging to note that all 23 spectroscopically confirmed
\TNSNe\ are correctly classified by SOFT (the lowest $\gamma_{\rm TN}$
classification grade is around 64\%).  This suggests a low incidence
of false negatives (mislabeling a \TNSN\ as a \CCSN), although this
should not be taken too far, as one would expect that the sub-sample
selected for spectroscopic follow-up includes the most easily
identified \TNSN\ candidates. Unfortunately the spectroscopic sample
cannot tell us anything about the rate of false positives (classifying
a \CCSN\ as a \TNSN).  The comparison of $z_{\rm SOFT}$ against
$z_{\rm spec}$ is shown in Figure~\ref{fig:zplot23}.  The mean $z_{\rm
  SOFT} - z_{\rm spec}$ residual is 0.024, and the rms scatter of the
residuals is 0.10, with a reduced $\chi^2$ statistic of 0.85.

The SOFT classification grades and redshift estimates for all 131
\TNSN\ candidates are summarized in Table~\ref{tab:ifa131}.  For each
SN in this table we have used the best available redshift measurement
to inform the prior.  The 23 spectroscopically confirmed SNe from
Table~\ref{tab:hiz23} appear again in Table~\ref{tab:ifa131}, now with
updated classification grades and improved parameter estimates due to
the addition of spectroscopic information.  Also included are redshift
priors derived from 14 likely SN host galaxies and photometric
redshift estimates for a further 22 possible hosts (see the final two
columns of Table~\ref{tab:ifa131}).  To account for possibile host
galaxy mismatches, the floor of the Gaussian redshift prior has been
raised to 10\% of the peak value, spread across the width of the
redshift parameter space.

\begin{deluxetable}{lllccc}
  \centering
  \tablewidth{0pt}
  \tablecaption{Redshift Validation Subset : 23 Spectroscopically Observed \TNSNe\label{tab:hiz23}}
  \tablehead{
    \colhead{Name} &
    \colhead{RA(J2000)} &
    \colhead{DEC(J2000)} &
    \colhead{$\gamma_{\rm TN}$} &
    \colhead{$z_{\rm spec}$} &
    \colhead{$z_{\rm SOFT}$\tablenotemark{a}}
  }
\startdata
SN2001fo  &  04:37:41.45  &  -01:29:33.1  &  1.000  &  0.772  &  0.716 $^{+0.033}_{-0.071}$ \\[0.3em]
SN2001fs  &  04:39:30.68  &  -01:28:21.9  &  0.928  &  0.874  &  1.169 $^{+0.085}_{-0.085}$ \\[0.3em]
SN2001hs  &  04:39:22.39  &  -01:32:51.4  &  0.998  &  0.833  &  0.946 $^{+0.137}_{-0.110}$ \\[0.3em]
SN2001hu  &  07:50:35.90  &  +09:58:14.2  &  0.991  &  0.882  &  1.048 $^{+0.085}_{-0.175}$ \\[0.3em]
SN2001hx  &  08:49:24.61  &  +44:02:22.4  &  0.999  &  0.799  &  0.899 $^{+0.123}_{-0.059}$ \\[0.3em]
SN2001hy  &  08:49:45.85  &  +44:15:31.8  &  0.998  &  0.812  &  0.886 $^{+0.149}_{-0.072}$ \\[0.3em]
SN2001iv  &  07:50:13.53  &  +10:17:10.4  &  1.000  &  0.397  &  0.410 $^{+0.019}_{-0.084}$ \\[0.3em]
SN2001iw  &  07:50:39.32  &  +10:20:19.1  &  1.000  &  0.340  &  0.391 $^{+0.030}_{-0.054}$ \\[0.3em]
SN2001ix  &  10:52:18.92  &  +57:07:29.6  &  0.921  &  0.711  &  0.642 $^{+0.098}_{-0.059}$ \\[0.3em]
SN2001iy  &  10:52:24.28  &  +57:16:36.1  &  1.000  &  0.568  &  0.591 $^{+0.175}_{-0.045}$ \\[0.3em]
SN2001jb  &  02:26:33.31  &  +00:25:35.0  &  0.936  &  0.698  &  0.811 $^{+0.111}_{-0.084}$ \\[0.3em]
SN2001jf  &  02:28:07.13  &  +00:26:45.1  &  0.881  &  0.815  &  0.851 $^{+0.123}_{-0.072}$ \\[0.3em]
SN2001jh  &  02:29:00.29  &  +00:20:44.2  &  1.000  &  0.885  &  0.764 $^{+0.033}_{-0.072}$ \\[0.3em]
SN2001jm  &  04:39:13.82  &  -01:23:18.2  &  0.999  &  0.978  &  0.883 $^{+0.058}_{-0.072}$ \\[0.3em]
SN2001jn  &  04:40:12.00  &  -01:17:45.9  &  0.897  &  0.645  &  0.797 $^{+0.045}_{-0.189}$ \\[0.3em]
SN2001jp  &  08:46:31.40  &  +44:03:56.6  &  0.640  &  0.528  &  0.512 $^{+0.071}_{-0.046}$ \\[0.3em]
SN2001kd  &  07:50:31.24  &  +10:21:07.3  &  0.934  &  0.936  &  0.893 $^{+0.111}_{-0.059}$ \\[0.3em]
SN2002P   &  02:29:05.71  &  +00:47:20.1  &  0.772  &  0.719  &  0.573 $^{+0.137}_{-0.098}$ \\[0.3em]
SN2002W   &  08:47:54.42  &  +44:13:42.9  &  0.822  &  1.031  &  1.067 $^{+0.111}_{-0.149}$ \\[0.3em]
SN2002X   &  08:48:30.54  &  +44:15:35.3  &  0.939  &  0.859  &  1.012 $^{+0.071}_{-0.163}$ \\[0.3em]
SN2002aa  &  07:48:45.28  &  +10:18:00.8  &  0.926  &  0.946  &  0.826 $^{+0.123}_{-0.045}$ \\[0.3em]
SN2002ab  &  07:48:55.70  &  +10:06:06.3  &  0.993  &  0.423  &  0.471 $^{+0.033}_{-0.111}$ \\[0.3em]
SN2002ad  &  10:50:12.19  &  +57:31:11.6  &  0.996  &  0.514  &  0.484 $^{+0.163}_{-0.045}$
\enddata
\tablenotetext{a}{\footnotesize The $z$ estimate determined by SOFT without a spectroscopic redshift prior. }
\end{deluxetable}


\begin{deluxetable}{lllccccl}
  \centering
  \tablewidth{\textwidth}
  \tablecaption{SOFT results for 131 \TNSN\ Candidates\label{tab:ifa131}}  
  \tablehead{
    \colhead{Name} &
    \colhead{RA(J2000)} &
    \colhead{DEC(J2000)} &
    \colhead{$\gamma_{\rm TN}$} &
    \colhead{$z_{\rm SOFT}$\tablenotemark{a} } &
    \colhead{$z_{\rm prior}$\tablenotemark{b} } &
    \colhead{$z_{\rm prior}$ Source\tablenotemark{c} } &
  }
  
\startdata

SN2001fi  &  02:26:38.87 &  +00:26:15.1  & 1.00  &  0.33 $^{+0.01}_{-0.01}$ & 0.320 $\pm$ 0.001 & IAUC\,7745 \\[1em]
SN2001fk  &  02:28:04.70 &  +00:46:17.7  & 1.00  &  0.81 $^{+0.01}_{-0.01}$ & 0.804 $\pm$ 0.001 & DEEP2\\[1em]
SN2001fn  &  02:29:00.49 &  +00:42:21.6  & 0.69  &  0.76 $^{+0.02}_{-0.04}$ & 0.186 $\pm 0.01$ & IAUC\,7745 \\[1em]
SN2001fo  &  04:37:41.45 &  -01:29:33.1  & 1.00  &  0.77 $^{+0.01}_{-0.01}$ & 0.772 $\pm$ 0.001 & B04\\[1em]
SN2001fq  &  04:38:04.51 &  -01:23:49.0  & 1.00  &  0.94 $^{+0.18}_{-0.06}$ & 0.01$\pm 0.01$  & IAUC\,7745\\[1em]

f0230a  &  02:26:09.11 &  +00:34:56.9  & 1.00  &  0.77 $^{+0.01}_{-0.01}$ & 0.771 $\pm$ 0.001 & DEEP2\\[1em]
f0230b  &  02:27:26.13 &  +00:48:27.7  & 0.53  &  0.24 $^{+0.00}_{-0.02}$ & \nodata & \\[1em]
f0230c  &  02:27:54.90 &  +00:31:48.4  & 1.00  &  1.00 $^{+0.06}_{-0.10}$ & \nodata & \\[1em]
f0230e  &  02:28:21.92 &  +00:24:26.8  & 0.60  &  1.02 $^{+0.01}_{-0.01}$ & 1.014 $\pm$ 0.001 & DEEP2\\[1em]
f0230f  &  02:29:35.86 &  +00:38:45.2  & 0.05  &  0.86 $^{+0.08}_{-0.06}$ & \nodata & \\[1em]

\enddata
\vspace{-2mm}
\tablenotetext{*}{~The complete table is available in the online version.}
\tablenotetext{a}{~Redshift determined with SOFT using the best available $z$ prior}
\tablenotetext{b}{~Independent SN or host galaxy redshift measurement, where available.}
\tablenotetext{c}{~Source of the redshift prior:\\
  {\em IAUC} = Spectroscopic $z$ measurement reported in IAU Circulars;\\
  {\em DEEP2} = Spectrocopic $z$ from the DEEP2 Redshift Survey, Data Release 2  (\url{http://deep.berkeley.edu/DR2});\\
  {\em SDSS} = Photo-$z$ from the SDSS Data Release 6 {\tt photoz2} catalog \citep{Oyaizu:2008}. }

\end{deluxetable}

The majority of these 131 candidates are classified by SOFT as \TNSNe:
90 objects have a \TNSN\ classification grade above 50\%, with 63 of
those above 90\%.  This is, however, somewhat lower than the 98
objects identified as \TNSNe\ in the
\citetalias{Barris:2006}\ analysis.  As will be explained below, we
include all 131 objects in our analysis, using the classification
grades to account for uncertainty in sample selection.  The mean
redshift uncertainty determined by SOFT for the complete set is 0.055,
broadly consistent with the scatter observed in
Figure~\ref{fig:zplot23} and seen in the SOFT validation tests using
SNe from the Sloan Digital Sky Survey (SDSS) and the Supernova Legacy
Survey (SNLS) \citep{Rodney:2010}.

A handful of SN candidates in Table~\ref{tab:ifa131} show
discrepancies between the redshift prior and the redshift estimate
determined by SOFT.  Three objects (f0848d, f0848f, f0848h) have
marginally significant ($<2\sigma$) differences between the assumed
host galaxy redshift and the SOFT redshift.  These host galaxy
redshift priors are taken from the SDSS photometric redshift catalog
\citep{Oyaizu:2008} and have very large uncertainties
($\sigma_z\ge 0.1$), so we accept the SOFT distribution as a
reasonable estimate.  Two other objects (SN2001ft, f0848b) have SOFT
redshifts that are significantly {\em lower} than the input redshift
priors.  Both of these objects have a classification grade
of $\gamma_{\rm TN}$=0, indicating that SOFT strongly classifies them
as \CCSNe.  In such cases the SOFT redshift estimation is very poor,
because \CCSNe\ are an inhomogeneous population that is only sparsely
sampled by the SOFT template library.  Our $\gamma_{\rm TN}$-weighted
counting procedure (described in Section~\ref{sec:DetectedCount})
ensures that these objects provide only a negligible contribution to
the count of detected \TNSNe.  Three other objects
(f0749t, f0848r, SN2001fn) have SOFT redshift estimates that are
significantly {\em higher} than their priors. The priors for the first
two are again from SDSS galaxies, and in this case both galaxies have
$>5\arcsec$ separation from the position of the SN.  We assume that
these are foreground galaxies, and the actual hosts have a low surface
brightness and are undetected by SDSS.  The final object, SN2001fn, is
well aligned with a galaxy in the SDSS that has a photometric redshift
of 0.7$\pm$0.3, although the reported spectroscopic redshift of the host galaxy
is 0.186.  Even with a strong spectroscopic prior at z=0.186, SOFT
finds a good classification for this object as a \TNSN\ at a redshift
of 0.76$^{+0.02}_{-0.04}$ -- in good agreement with the SDSS photo-z.
Without conclusive evidence in either direction, we adopt the SOFT
redshift estimate, noting that a change in the redshift of this single
object does not significantly affect our final rate measurements.

\section{The Revised IfA Deep \TNSN\ Rate}

The fundamental measurement needed to compute the \TNSN\ rate as a
function of redshift is simply a count of the number of detected SNe
as a function of redshift: $N_{\rm det}(z)$.  This count is computed
in Section \ref{sec:DetectedCount} for 9 redshift
intervals reaching out to z=1.1.  To convert this to a volumetric
rate, the detection count at each redshift is divided by the {\em
  control count}, $N_{\rm ctrl}(z) = {\rm SNuVol} \times V(z) \times
t_{\rm ctrl}(z)$. Here $V(z)$ is the volume of the survey in the
redshift range from $z$ to $z + dz$ and $t_{\rm ctrl}(z)$ is the effective survey period or {\em control time}: the total time span in which
a \TNSN\ exploding at redshift $z$ would be detectable by this survey.
Thus, the control count gives the number of SN detections that would be
expected from this survey if the SN rate per unit volume were constant
at all redshifts, fixed at 1 SNuVol=\SNuV.  Simulations to compute
$N_{\rm ctrl}(z)$ are described in Section \ref{sec:ControlCount}.

\subsection{The Detected Count}
\label{sec:DetectedCount}

\begin{figure*}[tb!]
  \centering
  \ifrgb{
    \includegraphics[draft=False,width=0.48\textwidth]{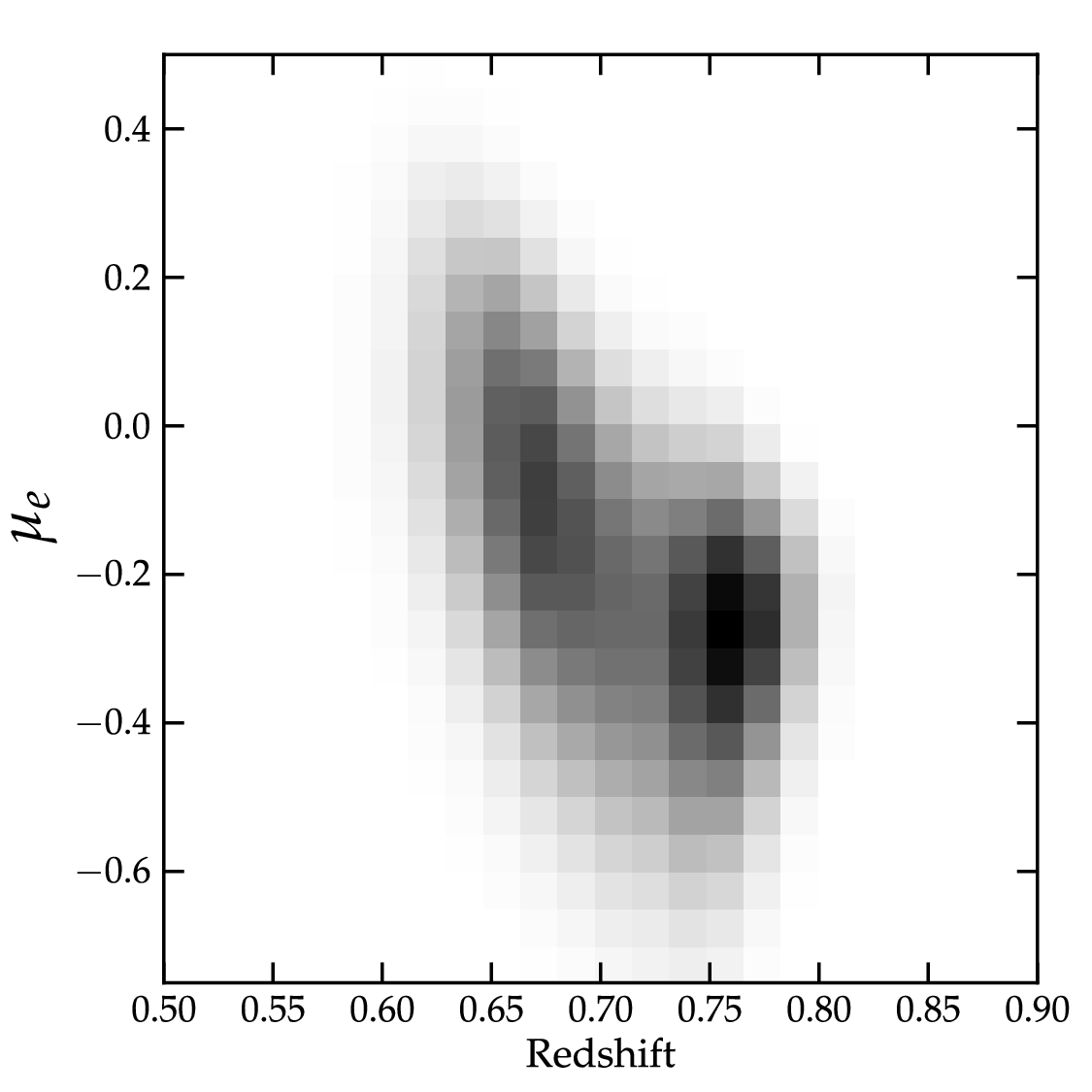} 
    \includegraphics[draft=False,width=0.48\textwidth]{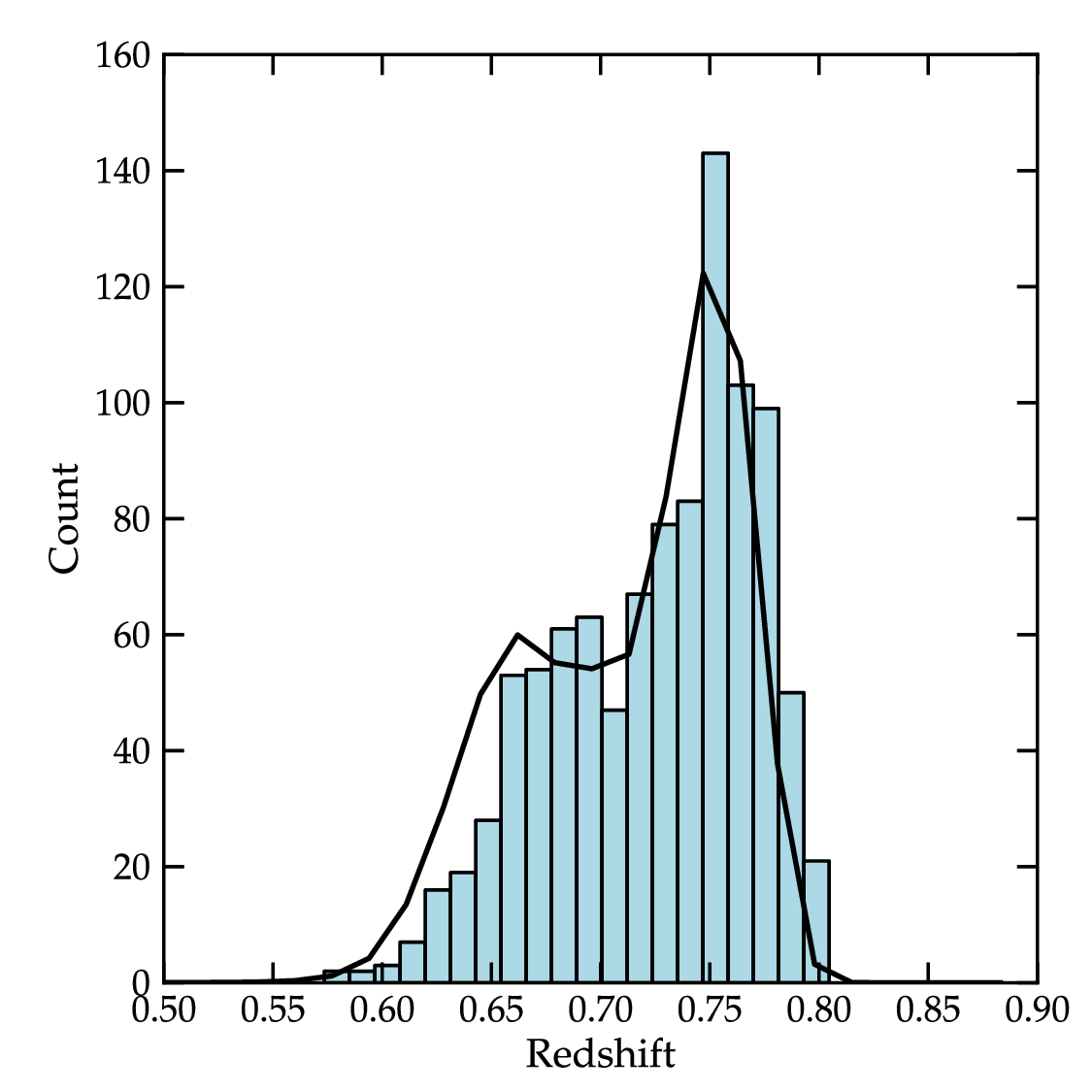} 
  }\else{
    \includegraphics[draft=False,width=0.48\textwidth]{f02a_gray} 
    \includegraphics[draft=False,width=0.48\textwidth]{f02b_gray} 
  } \fi
  \caption[MCMC counting from SOFT redshift distributions]{ To tally
    the contribution from each individual SN candidate over a series
    of redshift bins, we utilize a Markov Chain Monte Carlo series,
    drawing 1000 random redshifts that reproduce the redshift
    distribution measured by SOFT.  {\bf (a)} The composite fuzzy
    membership function (equivalent of a posterior probability
    distribution) for SN\,f0230a in the (z,\mue) plane.  {\bf (b)}
    After marginalizing over \mue, the distribution in redshift (solid
    line) is used as the target distribution for our MCMC series.  The
    blue histogram shows the results of 1000 MCMC draws, effectively
    reproducing the highly irregular probability distribution. }
  \label{fig:zMonteCarlo}
\end{figure*}

In the absence of a strong spectroscopic redshift prior, the redshift
distributions produced by SOFT are typically broad and asymmetric,
sometimes with multiple peaks - making it impossible to pin them down
into any single redshift bin for counting. To incorporate these
complex uncertainties into our \TNSN\ detection counts, a Markov Chain
Monte Carlo (MCMC) sampling test was executed for each SN, as
illustrated in Figure~\ref{fig:zMonteCarlo}.  The MCMC provides 1000
redshift realizations for each SN, distributed so as to reconstruct
the measured SOFT redshift distribution.

After repeating this for all candidates, we have 1000 sets of 131
redshift assignments, with each set representing a possible redshift
distribution for the entire IFA Deep SN sample.  To count the number
of detected SNe, each set of 131 redshifts is binned into intervals of
width $\Delta z$=0.1 from 0.0 to 1.4. To incorporate the uncertainty
of classification into our counts, we allow each SN to contribute a
fractional count to the redshift bin that it falls in for a given
simulation.  This fractional count is drawn from a Gaussian
distribution centered on the composite \TNSN\ classification grade,
$\gamma_{\rm TN}$, with a width of $\sigma=0.05$. This 5\% scatter
reflects the uncertainty of the classification grades deduced from
testing SOFT against validation data in \citet{Rodney:2009}.  With
each SN contributing a count of less than unity, the average of the
total count across all redshifts is $\sum\gamma_{\rm TN}=84.5$, rather
than 131.

After binning up all 1000 Monte Carlo realizations, we have 1000
measurements for the count in each bin.  Taking the mean of this
1000-element vector gives us our central count estimate for that
redshift interval, and the standard deviation provides an estimate of
the statistical uncertainty that accounts for error in both the
classification and the redshift estimation.  The counts of
detected SNe in each redshift bin are given in
Table~\ref{tab:newRates}a, along with the statistical and systematic
uncertainties (described in
Section~\ref{sec:SystematicUncertaintyEstimates}).

\begin{deluxetable}{c | r l l l | r l | r l l}
  \centering
  \tablewidth{0pt}
  \tablecaption{New Determination of IfA Deep \TNSN\ Rates\label{tab:newRates}}
  \tablecolumns{10}

  \tablehead{
    \multirow{2}{*}{$z$~\tablenotemark{a}} & 
    \multicolumn{4}{c}{Detected Count~\tablenotemark{b}} & 
    \multicolumn{2}{c}{Control Count~\tablenotemark{c}} & 
    \multicolumn{3}{c}{SN Rate~\tablenotemark{d}} \\
    & \colhead{$N_{\rm det}$} & \colhead{$\delta N_{z}$} & \colhead{$\delta N_{\rm Poiss}$} & \colhead{$(\delta N_{\rm syst})$} &
    \colhead{$N_{\rm ctrl}$} & \colhead{$(\delta N_{\rm syst})$} &
    \colhead{SNR}  & \colhead{$\delta$SNR$_{stat}$} & \colhead{($\delta$SNR$_{syst}$)}
  }
  \startdata
 0.15  & 1.95    &  $\pm$0.12 &  $\pm$1.40 & $\left(^{+0.04}_{-0.32}\right)$  &  6.1 & $\left(^{+0.6}_{-1.3}\right)$ & 0.32  & $\pm$0.23 &  $\left(^{+0.07}_{-0.06}\right)$ \\[1em]
 0.35  & 4.01    &  $\pm$0.91 &  $\pm$2.00 & $\left(^{+0.29}_{-0.00}\right)$  & 11.7 & $\left(^{+1.2}_{-2.2}\right)$ & 0.34  & $\pm$0.19 &  $\left(^{+0.07}_{-0.03}\right)$ \\[1em]
 0.45  & 5.11    &  $\pm$1.14 &  $\pm$2.26 & $\left(^{+1.74}_{-0.16}\right)$  & 16.6 & $\left(^{+2.2}_{-3.0}\right)$ & 0.31  & $\pm$0.15 &  $\left(^{+0.12}_{-0.04}\right)$ \\[1em]
 0.55  & 6.49    &  $\pm$1.42 &  $\pm$2.55 & $\left(^{+0.17}_{-1.30}\right)$  & 20.4 & $\left(^{+2.3}_{-4.3}\right)$ & 0.32  & $\pm$0.14 &  $\left(^{+0.07}_{-0.07}\right)$ \\[1em]
 0.65  & 10.09   &  $\pm$1.75 &  $\pm$3.18 & $\left(^{+0.95}_{-1.09}\right)$  & 20.8 & $\left(^{+2.5}_{-5.5}\right)$ & 0.49  & $\pm$0.17 &  $\left(^{+0.14}_{-0.08}\right)$ \\[1em]
 0.75  & 14.29   &  $\pm$2.19 &  $\pm$3.78 & $\left(^{+0.60}_{-2.03}\right)$  & 20.9 & $\left(^{+3.0}_{-7.0}\right)$ & 0.68  & $\pm$0.21 &  $\left(^{+0.23}_{-0.14}\right)$ \\[1em]
 0.85  & 15.43   &  $\pm$2.09 &  $\pm$3.93 & $\left(^{+1.78}_{-0.00}\right)$  & 19.9 & $\left(^{+4.1}_{-7.7}\right)$ & 0.78  & $\pm$0.22 &  $\left(^{+0.31}_{-0.16}\right)$ \\[1em]
 0.95  & 13.21   &  $\pm$2.31 &  $\pm$3.63 & $\left(^{+0.32}_{-1.43}\right)$  & 17.3 & $\left(^{+5.7}_{-7.1}\right)$ & 0.76  & $\pm$0.25 &  $\left(^{+0.32}_{-0.26}\right)$ \\[1em]
 1.05  & 11.01   &  $\pm$2.08 &  $\pm$3.32 & $\left(^{+1.28}_{-1.83}\right)$  & 13.9 & $\left(^{+6.8}_{-6.1}\right)$ & 0.79  & $\pm$0.28 &  $\left(^{+0.36}_{-0.41}\right)$ \\[1em]
  \enddata
  \tablenotetext{a}{ \footnotesize Central redshift for bins of width 0.1 (except
    for the first bin, centered on $z=0.15$, which has width 0.3) }
  \tablenotetext{b}{ \footnotesize $N_{\rm det}$ gives the number of detected
    \TNSNe\ in each bin, and $\delta N_{z}$ is the uncertainty due to
    the redshift distributions, described in
    Section~\ref{sec:DetectedCount}. The Poisson uncertainty $\delta
    N_{\rm Poiss}$ is simply $\sqrt{ N_{\rm det}}$, and the systematic
    uncertainty $\delta N_{\rm syst}$ is taken from the
    Table~\ref{tab:IFAsystematics}a. }
  \tablenotetext{c}{ \footnotesize $N_{\rm pred}$ gives the control count of
    \TNSNe\ as derived in Section~\ref{sec:ControlCount}.  The
    systematic uncertainty $\delta N_{\rm syst}$ is taken from
    Table~\ref{tab:IFAsystematics}b. }
  \tablenotetext{d}{ \footnotesize The derived \TNSN\ rate measurements in units of
    \SNuV, computed as SNR$=N_{\rm det} / N_{\rm ctrl}$, with
    accompanying statistical ($\delta$SNR$_{stat}$) and systematic
    ($\delta$SNR$_{syst}$) uncertainties. }
\end{deluxetable}

\subsection{The Control Count}
\label{sec:ControlCount}

The second component of the \TNSN\ rate calculation, $N_{\rm ctrl}$,
is derived from a Monte Carlo simulation of $\sim$5000 SNe that
replicates the IfA Deep survey observations.  The simulation is
defined over a grid of redshift values from $z=0.01$ to 1.6 in steps
of $\delta z=0.02$.  At each point on the grid we compute $N_{\rm
  exp}(z)$, the total number of SNe that explode each year within the
IfA Deep survey area in a shell of thickness $\delta z$.  This number
is calculated as $ N_{\rm exp}(z) = 1~{\rm SNuVol} \times V(z) \times
1~{\rm yr}$. Here $V(z)$ is the volume of a shell at redshift $z$,
using a total survey area of 2.29 square degrees and assuming a
\LCDM\ cosmology.\footnote{ A flat, $\Lambda$-dominated universe with
  H$_0$=71, \Om=0.26, \Ol=0.74 \citep{Komatsu:2010}.}

Having computed the total number of explosions for each point on the
redshift grid, we now want to simulate how these SNe would be
distributed in magnitude space.  We assume a luminosity function for
the \TNSN\ population composed of three Gaussian distributions
representing the overluminous objects (Ia$+$ or SN 1991T-like), which
make up 20\% of the population, normal Ia's at 64\%, and underluminous
\TNSNe\ (Ia$-$ or SN 1991bg-like) comprising 16\% \citep{Li:2001}.
The Gaussians have $\sigma=$0.45, 0.30, and 0.50 magnitudes for the
Ia$+$, Ia, and Ia$-$ components, respectively. The same luminosity
function is used for all redshift bins.  

Each of the $\sim$5000 simulated SNe now has an assumed redshift $z$
and a peak magnitude $m_{\rm pk}$ drawn from the luminosity functions.
These parameters are then applied to construct a model light curve
using the apparatus of the SOFT template library \citep{Rodney:2009}.
The simulated light curve includes cosmological dimming appropriate
for the redshift by assuming a \LCDM\ cosmology, as well as extinction
due to host galaxy dust that is drawn from a ``galactic line of
sight'' distribution \citep{Hatano:1998} as implemented in
\citet{Tonry:2003} and \citet{Wood-Vasey:2007}.  Briefly, we draw a
host extinction value for each simulated SN from a probability
distribution constructed by assuming that 25\% of hosts are bulge
systems and 75\% are disk systems. The extinction distribution curves
of \citet{Hatano:1998} are approximated by $f(A_B)\propto
0.02\delta(A_B)+10^{-1.25-A_B}$ for bulge systems and $f(A_B)\propto
0.02\delta(A_B)+e^{-3.77-A_B^2}$ for disk systems.

The control time $t_{ctrl}(z,m_{\rm pk})$ for this model light curve
is calculated by counting the days on which this SN could explode and
subsequently be detected by the IfA Deep survey.  To determine if a
model SN would be detected, we count up the number of IfA Deep
observation epochs for which the synthetic light curve is brighter
than the detection threshold.\footnote{ As was done for the actual
  survey, the I band is the only search filter, and the detection
  limit is set at the single-observation 5-$\sigma$ point source limit
  of I=25.2 mags.}  If the tally of detection epochs is $\ge 5$ then
we consider this SN to be detected.

Multiplying together the number of SN explosions $N_{exp}(z,m_{pk})$
per year and the control time $t_{ctrl}(z,m_{pk})$ gives us a final
estimate for the control count of SN detections
$N_{ctrl}(z,m_{pk})$. Integrating over all possible peak magnitudes
gives us the final control count of \TNSN\ detections in each redshift
bin, presented in Table~\ref{tab:newRates}b.  The systematic
uncertainty estimates included in Table~\ref{tab:newRates} are
derived in the following section.

\section{Systematic Uncertainty Estimates}
\label{sec:SystematicUncertaintyEstimates}

Both of the principle components of our rate calculation -- the
detection count and the control count -- can potentially introduce
systematic errors into the SN rate values.  A suite of systematic
uncertainty tests were performed in which a parameter of interest was
adjusted, the classification and redshift estimation were repeated for
all candidates, and new rate estimates were measured.  The difference
between the revised rate and our baseline rate measurement was taken
as the systematic error estimate, separately for each redshift bin.
All tests are described below, and the resulting uncertainty estimates
are reported in Table~\ref{tab:IFAsystematics}.

\subsection{Systematic Errors in $N_{det}$}

{\bf Peculiar SNe:} If a normal \TNSN\ gets matched by a highly
  peculiar SN template we could derive an erroneous redshift. We don't
  expect a significant number of peculiar Ia's in a survey of this
  size.  This systematic test removes the very unusual SN~2000cx and
  SN2002cx from the SOFT template library.

{\bf Host Galaxy Dust:}
First, we change the ratio of total to selective extinction to \Rv=2.1, accommodating the possibility that dust around
extragalactic SNe is very different from the Milky Way extinction law
of \Rv={3.1}
\citep{Astier:2006,Kowalski:2008,Kessler:2009,Poznanski:2009}.
Next, we change the prior on the extinction parameter \Av\ from
the ``Galactic Line of Sight'' prior to an exponential distribution
$p(\Av)\propto \exp(-\Av/5)$ that approximates the extreme inclination
models of \citet{Riello:2005}, following \citet{Neill:2006}.  

{\bf Redshift Prior:} For objects without a spectroscopic redshift
measurement or a host galaxy photo-$z$, we have relied on a redshift
prior that scales with the cosmological volume element, going as
$z^2$.  This can introduce a systematic bias towards higher $z$
estimates.  An alternative and differently biased redshift prior can
be derived from the control count, as given in
Table~\ref{tab:newRates}.  The control count vs. z for this survey is
well approximated by a Gaussian with $\mu=0.7$ and $\sigma=0.38$,
so we estimate the systematic uncertainty by using this function for
the redshift prior.

{\bf ``Fuzzy AND'' Rejection Threshold}
As detailed in \citet{Rodney:2010}, the SOFT redshift estimate for any
candidate SN is derived by combining the fuzzy membership functions
from multiple templates using the Fuzzy AND operation.  To prevent the
combination from being dominated by extreme outliers, we have used a
rejection threshold of 5\% on the integrated membership grade
$\gamma_{TN}$.  To test whether this value is introducing systematic
effects, we reprocessed the SN light curves using threshold values of
3\% and 10\%.

\begin{deluxetable}{l lllll llll}
  \tablewidth{0pt}
  \tabletypesize{\small}
  \tablecaption{Systematic Uncertainty Estimates}
  \tablecolumns{10}
  \tablehead{ 
  & \multicolumn{9}{c}{(a) Change in Detected Count}\\[1em]
  \multicolumn{1}{r}{Redshift:} &  \multicolumn{1}{c}{0.25} & \multicolumn{1}{c}{0.35} & \multicolumn{1}{c}{0.45} & \multicolumn{1}{c}{0.55} & 
             \multicolumn{1}{c}{0.65} & \multicolumn{1}{c}{0.75} &  \multicolumn{1}{c}{0.85} & \multicolumn{1}{c}{0.95} & \multicolumn{1}{c}{1.05}
  }
  \startdata

  drop peculiar templates \dotfill         & $-0.00$ & $+0.02$ & $+0.72$ & $+0.00$ & $-0.69$           & $+0.22$        & $+0.06$  & $-1.33$           & $+0.93$ \\[2em]
  change $R_V$ by 1 \dotfill               & $-0.32$ & $+0.14$ & $+1.42$ & $+0.17$ & $-0.72$           & $-0.89$        & $+0.08$  & $-0.34$           & $+0.20$ \\[2em]
  flat \Av\ prior \dotfill                 & $+0.04$ & $+0.20$ & $+0.71$ & $-0.32$ & $+0.00$           & $-0.50$        & $+0.71$  & $-0.39$           & $-0.77$ \\[2em]
  Gaussian $z$ prior \dotfill              & $+0.01$ & $+0.10$ & $-0.06$ & $-0.08$ & $+0.34$           & $-0.52$        & $+0.13$  & $-0.02$           & $-0.02$ \\[2em]
  modified fuzzy AND threshold \dotfill    & $+0.00$ & $+0.11$ & $-0.15$ & $-1.26$ & $^{+0.89}_{-0.46}$ & $^{+0.56}_{-1.67}$  & $+1.63$  & $^{+0.32}_{-0.04}$ &  $^{+0.85}_{-1.66}$ \\[2em]
  \multirow{2}{*}{{\bf TOTAL :}}           & $+0.04$ & $+0.29$ & $+1.74$ & $+0.17$ & $+0.95$           & $+0.60$        & $+1.78$  & $+0.32$           & $+1.28$ \\
                                           & $-0.32$ & $-0.00$ & $-0.16$ & $-1.30$ & $-1.09$           & $-2.03$        & $-0.00$  & $-1.43$           & $-1.83$ \\[2em]

  \tableline\\
  & \multicolumn{9}{c}{(b) Change in Control Count}\\[1em]
  \multicolumn{1}{r}{Redshift:} &  \multicolumn{1}{c}{0.25} & \multicolumn{1}{c}{0.35} & \multicolumn{1}{c}{0.45} & \multicolumn{1}{c}{0.55} & 
             \multicolumn{1}{c}{0.65} & \multicolumn{1}{c}{0.75} &  \multicolumn{1}{c}{0.85} & \multicolumn{1}{c}{0.95} & \multicolumn{1}{c}{1.05}\\
  \tableline\\

  \multirow{2}{*}{Alternate SN luminosity function} \dotfill & $+0.2$ & $+0.4$ & $+1.0$ & $+0.4$ & $+0.5$ & $+0.0$ & $+0.0$ & $+0.0$ & $+0.3$ \\
                                                             & $-0.3$ & $-0.5$ & $-0.5$ & $-1.1$ & $-0.8$ & $-0.7$ & $-1.2$ & $-1.1$ & $-1.5$ \\[2em]

  \multirow{2}{*}{Minimum no. detections  $\pm$ 1}  \dotfill & $+0.55$ & $+1.1$ & $+1.95$ & $+2.3$ & $+2.45$ & $+3.05$ & $+4.1$ & $+5.65$ & $+6.8$ \\
                                                             & $-1.05$ & $-1.8$ & $-2.25$ & $-2.6$ & $-3.15$ & $-4.45$ & $-5.6$ & $-5.7$  & $-5.15$ \\[2em]

  \multirow{2}{*}{Filling factor $\pm$ 4\% }        \dotfill & $+0.00$ & $+0.05$ & $+0.15$ & $+0.0$ & $+0.0$  & $+0.0$  & $+0.0$  & $+0.0$  & $+0.0$ \\
                                                             & $-0.35$ & $-0.55$ & $-0.75$ & $-1.5$ & $-1.45$ & $-1.85$ & $-2.05$ & $-1.65$ & $-1.35$ \\[2em]

  Exponential dust profile  \dotfill & $-0.65$ & $-1.1$ & $-1.7$ & $-2.9$ & $-4.2$ & $-5.1$ & $-4.8$ & $-3.8$ & $-2.5$ \\

  \tableline\\
  \multirow{2}{*}{{\bf TOTAL :}}           &  $+0.59$ & $+1.17$ & $+2.20$ & $+2.33$ & $+2.50$ & $+3.05$ & $+4.10$ & $+5.65$ & $+6.81$ \\
                                           &  $-1.32$ & $-2.24$ & $-2.96$ & $-4.32$ & $-5.47$ & $-7.02$ & $-7.75$ & $-7.13$ & $-6.07$ \\

  \enddata
  \label{tab:IFAsystematics}
\end{deluxetable}

\begin{figure*}[tb!]
  \centering
  \ifrgb{
    \includegraphics[width=\textwidth]{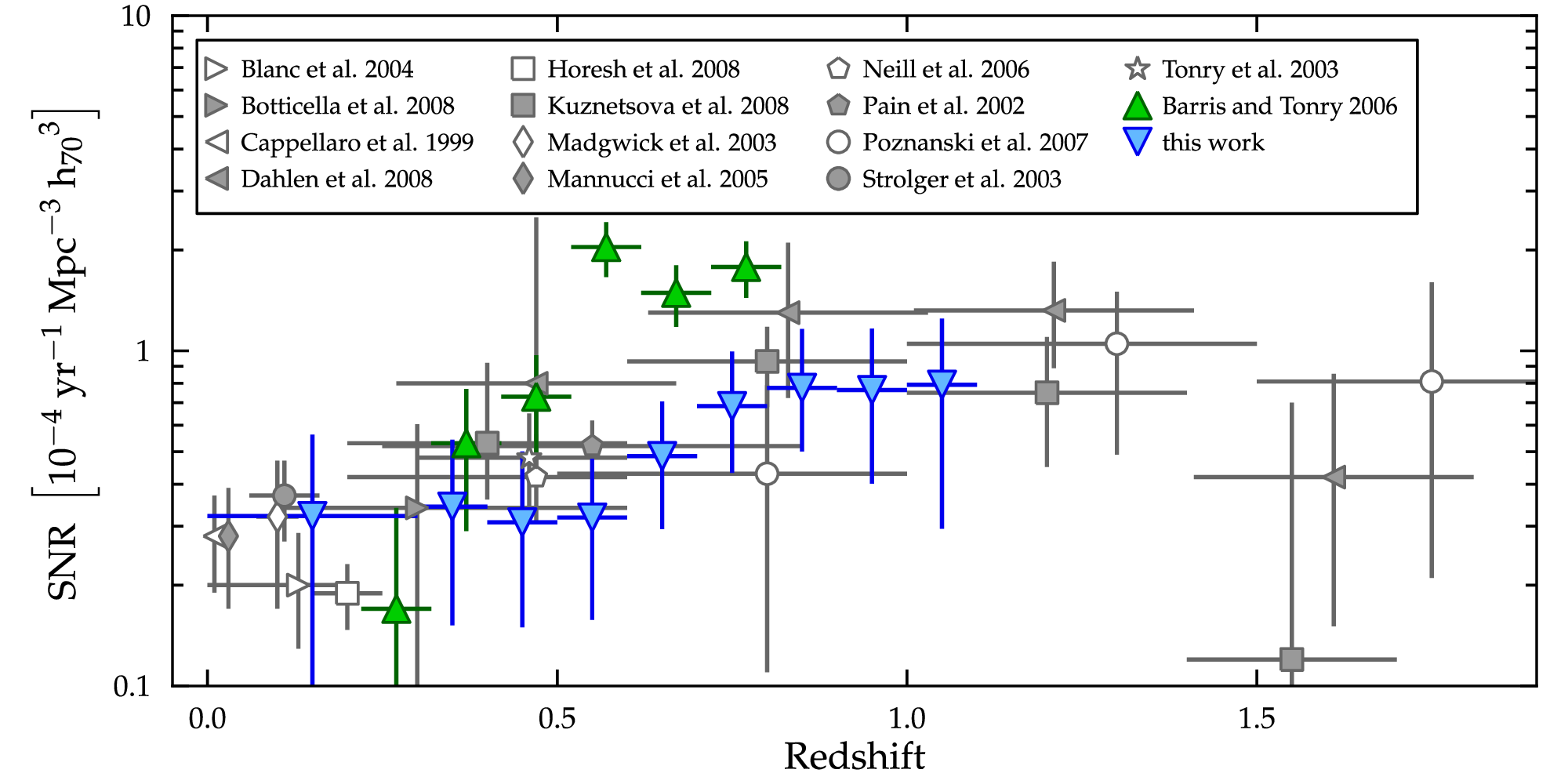} 
  }\else{
    \includegraphics[width=\textwidth]{f03_gray} 
  } \fi
  \caption[New \TNSN\ Rates from IFA Deep]{ Results of the final
    \TNSN\ rate calculation.  The original IFA Deep rates
    \citepalias{Barris:2006} are shown as upward-pointing green
    triangles. The new rate measurements described here are plotted as
    downward-pointing blue triangles.  Error bars on the new rate
    points include both statistical and systematic uncertainties.  A
    compilation of rates from the literature are plotted as gray
    symbols, with references as given in the legend.  }
  \label{fig:IFArates}
\end{figure*}
\nocite{Blanc:2004}
\nocite{Botticella:2008}
\nocite{Cappellaro:1999}
\nocite{Dahlen:2008}
\nocite{Horesh:2008}
\nocite{Kuznetsova:2008}
\nocite{Madgwick:2003}
\nocite{Mannucci:2005}
\nocite{Neill:2006}
\nocite{Pain:2002}
\nocite{Poznanski:2007}
\nocite{Strolger:2003}
\nocite{Tonry:2003}

\subsection{Systematic Errors in $N_{\rm ctrl}$}

{\bf \TNSN\ Luminosity Function:}
There is substantial uncertainty in the balance of the SN luminosity
function between the overluminous Ia$+$, the normal Ia, and the
underluminous Ia$-$ sub-classes, provided by \citet{Li:2001}. 
We first modify the baseline (20\%, 64\%, 16\%) to (13\%, 77\%, 10\%)
and then to (27\%, 51\%, 22\%).  

{\bf Detection Thresholds:}
The actual process of extracting 131 likely \TNSN\ candidates from
among the many thousands of potential transient objects in the IfA
Deep survey was quite complex (for details see \citet{Barris:2004b}).
Our proxy for the selection process used the simple requirement that a
SN must be above the 5-$\sigma$ point source threshold in 5 epochs in
order to be counted.  Our systematic test modifies this to 4 and then
6 epochs.  The reported systematic error estimate is half of the
difference between the baseline and the modified control count.

{\bf Survey Area:}
The computation of the survey area was greatly simplified in our Monte
Carlo procedure by assuming a 90\% filling factor for volume
calculations. We recomputed the control count using alternate
filling factors of 86\% and 94\%.

{\bf Host Galaxy Dust:}
As in the detected count systematic uncertainty test above, we modify
the host galaxy dust first to use $R_V=2.1$ and then to follow an
exponential $A_V$ distribution.

\section{Results} 

A summary of the principal components of the SN Rate calculation is
given in Table~\ref{tab:newRates}.  The final tally of the number of
detected \TNSNe\ is given in column 2, with statistical and systematic
uncertainties in columns 3--5.  The control count of \TNSNe\ is given
in column 6, followed by systematic uncertainty estimates.  The
resulting SN Rate numbers and associated uncertainties appear in the
final three columns.

These revised rate estimates are significantly different from the
numbers derived in \citetalias{Barris:2006}.  These differences arise from improvements in
both the measurement of $N_{det}$ and the calculation of
$N_{ctrl}$. With an improved SN light curve analysis program in SOFT,
we now have more robust photometric redshift estimates and
classification grades for all 131 SNe.  SOFT also has new access to
more detailed prior information with the addition of 36 host galaxy
redshift measurements from spectra and photo-z's. These improvements
have allowed us to reliably extend the rate estimate beyond the z=0.75
limit applied by \citetalias{Barris:2006}, reaching now out to a mean redshift of 1.05 in
the final bin.

Using an updated Monte Carlo simulation, we have generated a revised
control count for the number of SNe detectable by the IfA Deep survey
as a function of redshift.  Our simulations consistently disagree with
\citetalias{Barris:2006}\ in that we compute a larger control count in every one of the
redshift bins.  However, we have also performed a series of systematic
uncertainty tests, and we find that when systematic errors are
included, our control counts are consistent with those from \citetalias{Barris:2006}.

In Figure~\ref{fig:IFArates} we show the final rate measurements from
the IfA Deep survey, compared to other rates from the literature.  The
original IfA Deep rates are shown in green, and the newly derived
rates are plotted in blue.  A compilation of rates from the literature
is plotted in the background in gray.  The new IfA rate measures are
clearly in much better agreement with the consensus of literature
rates than were the \citetalias{Barris:2006}\ values.  Our new IfA results indicate that
the \TNSN\ rate increases with redshift out to at least z=1.0, with no
evidence of reaching a peak in the redshift range sampled.


\bibliographystyle{apj}
\bibliography{bibdesk}


\end{document}